# PACER

## Peripheral Activity Completion Estimation and Recognition

Dr. Daniel Ross Moore
Center for Efficient, Scalable and Reliable Computing
Dept. of Electrical and Computer Engineering
North Carolina State University, Raleigh, USA
drmoore2@ncsu.edu

Dr. Alexander G. Dean
Center for Efficient, Scalable and Reliable Computing
Dept. of Electrical and Computer Engineering
North Carolina State University, Raleigh, USA
agdean@ncsu.edu

*Abstract*—Embedded peripheral devices such as memories, sensors and communications interfaces are used to perform a function external to a host microcontroller. The device manufacturer typically specifies worst-case current consumption and latency estimates for each of these peripheral actions. Peripheral Activity Completion, Estimation and Recognition (PACER) is introduced as a suite of algorithms that can be applied to detect completed peripheral operations in real-time. By detecting activity completion, PACER enables the host to exploit slack between the worst-case estimate and the actual response time. These methods were tested independently and in conjunction with IODVS on multiple common peripheral devices. For the peripheral devices under test, the test fixture confirmed decreases in energy expenditures of up to 80% and latency reductions of up to 67%.

## I. Introduction

Embedded systems are often constrained by timing and energy budgets because both factors affect the resultant cost and size of the system. Peripheral devices external to the microcontroller (MCU) such as those shown in Figure 1 can play a significant role in system-wide energy consumption. There are many methods available for decreasing the static power usage of peripherals [1] [2] [3]. PACER decreases dynamic power consumption and latency by exploiting the slack between actual versus worst-case operation time.

Device manufacturers derive and specify the worst-case operation duration by summing exacerbating factors including age, temperature and voltage. Using the worst-case operation time as a naïve guideline, the worst-case energy consumption of a given operation is characterized by (1).

$$E_{op-wc} = \int_0^{t_{op}} P_{op}(t)dt + \int_{t_{op}}^{t_{slack}} P_{slack}(t)dt \quad (1)$$

Where $t_{op}$ and $P_{op}$ are the time and power comprising the actual operation while $t_{slack}$ and $P_{slack}$ are the time and power comprising the period between operation completion and the worst-case execution time.

Most peripheral devices provide a mechanism for signaling that operations completed earlier than the maximum. However, using these mechanisms results in sub-optimal power performance. For example, a common method of detecting write completion on external non-volatile memory relies on polling a status register. Performing this signaled method has power and energy consequences:

$$P_{overhead} = P_{MCU} + P_{MCD} + P_{Comm} + P_{Match} + P_{Dev} \quad (2)$$

- $P_{MCU}$: MCU must be active while polling
- $P_{MCD}$: MCU communications driver must be active
- $P_{Comm}$: Communications incurs $P = cfV_{dd}^2$ penalty
- $P_{Match}$: MCU and device voltages must be matched.
  - Neither can use dynamic voltage scaling
- $P_{Dev}$: Device communications driver must be active

$$E_{op-sig} = \int_0^{t_{op}} \big(P_{op}(t) + P_{overhead}(t)\big)dt \quad (3)$$

The components of $P_{overhead}$ are highly variable between microcontrollers, systems and devices. The signal may involve protocol-level communication or it may be as simple as an interrupt pin and that signal may traverse PCB traces with considerable capacitance. $E_{op-sig}$ can exceed $E_{op-wc}$.

Both interface methods incur a power penalty and the naïve worst-case method also incurs a latency penalty. As the energy cost of computation continues to decrease in modern microcontrollers, it becomes more rewarding to use onboard intelligence to minimize the impact of power and latency penalties. PACER develops adaptive timing, current usage and charge consumption heuristics for estimating or recognizing early completion of peripheral operations, thus reducing total latency and energy consumption.

The prediction is verified in real-time against the actual state and the heuristic is updated with the results. In this fashion, the algorithms are resistant to variations in behavior that may occur across the lifecycle of the device. PACER is evaluated against a variety of embedded peripherals and is shown to significantly decrease both energy consumption and latency of peripherals with minimal computational overhead.

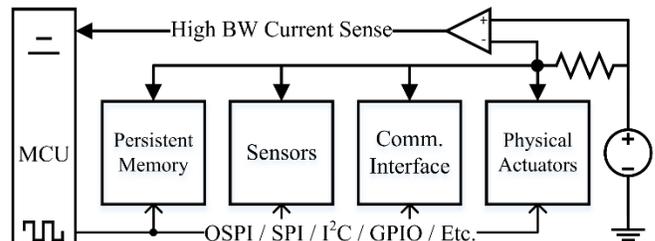

*Figure 1: Typical Embedded System with Device Current Feedback*

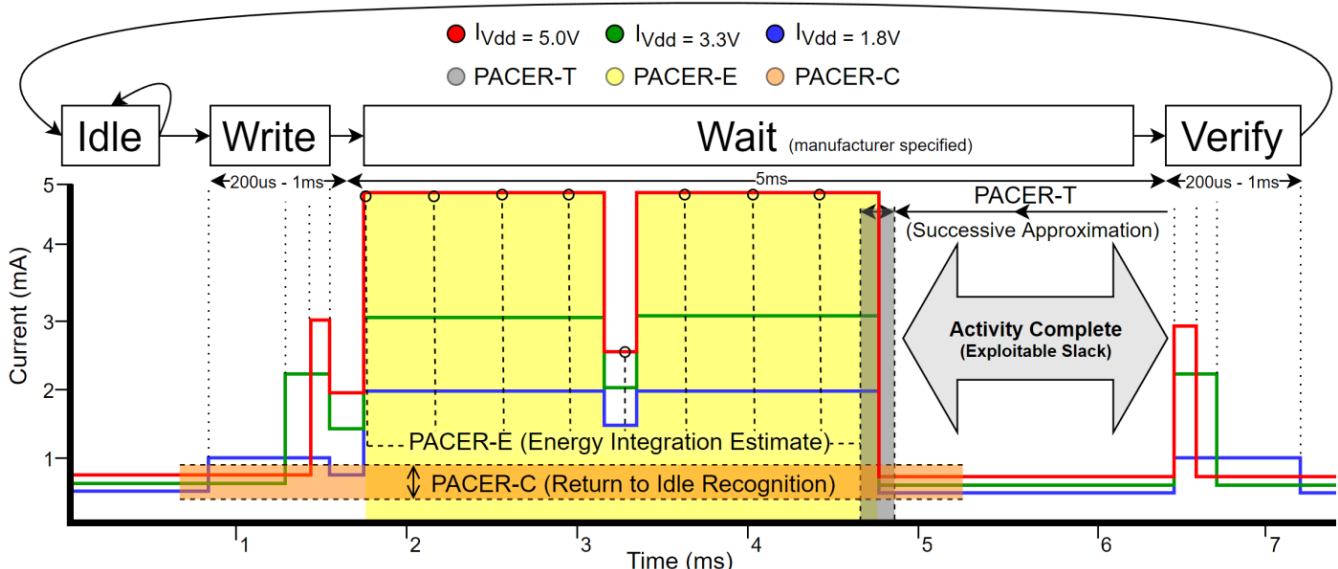

*Figure 2: A Typical External Memory Transaction with IODVS and PACER*

Figure 2 shows the current profile for the common peripheral operation of writing a page to EEPROM. The manufacturer-specified mandatory wait period is 5ms, beginning about the 1.75ms mark. As the device transitions through the Idle → Write → Wait → Verify states, it can be inferred from the current profile that the operation completed by the 4.75ms mark and that it was not necessary to delay until approximately 6.5ms per the specification. This 1.75ms differential is slack that can be exploited to decrease latency.

There are a wide variety of peripheral devices with a correspondingly wide variety of completion determinism and current profiles. PACER introduces three methods by which the host MCU can estimate or detect early completion of peripheral operations while also minimizing computational overhead. Devices with highly deterministic timing respond best to the timing heuristic while those with variable timing respond best to current or charge heuristics. Through low-overhead early completion detection, PACER is able to decrease both latency and system-wide energy consumption.

## II. RELATED WORK

Intra-Operation Dynamic Voltage Scaling [4] (IODVS) has been shown to significantly reduce the energy consumption of embedded peripherals (Flash, EEPROM, sensors, etc.) during their voltage-independent states. These states typically occur during mandatory delay periods while the peripheral completes a specified operation. When implementing IODVS, the host MCU and peripheral devices are placed on different voltage domains throughout the course of the voltage-independent state. Because of this, it is not possible for the MCU to poll the peripheral device for operation completion. Polling is also shown to be a rather costly operation due to (2) and (3). Without the ability to communicate to the peripheral device, PACER is necessary to achieve minimal operation latencies.

### A. Timing Heuristic

Peripheral operations can vary in their latency or completion times due to a number of factors. Temperature can significantly affect the completion time for peripherals with deterministic timing requirements such as DRAM [5]. Device aging can also affect timing due to a number of issues resulting from fundamental semiconductor physics [6]. Furthermore, some devices simply have non-deterministic completion times due to features such as MMUs and caches that are implemented in various data storage devices like Micro-SD cards, or age and wear as they effect FLASH storage timing.

Because the latency can vary significantly between operations, it is necessary to develop a timing heuristic that can adapt to slowly changing effects like age and temperature as well as rapidly changing factors like cache hits and misses. Adaptive delay estimation is not a new problem [7] and research continues to compensate for non-deterministic delay with different approaches for wireless communications, control systems and mass storage latency [8].

### B. Energy Heuristic

For devices with highly variable timing and dynamic current consumption characteristics, integrating the current consumption of the device throughout an operation can allow for better detection of completion. Some operations can be characterized by the amount of charge necessary to complete them. This technique is referred to as "coulomb counting" and is a common technique used to determine the state of charge in rechargeable batteries [9].

### C. Current Heuristic

The completion of some peripheral operations are easily detectable by their current consumption profile. These devices have a distinct and deterministic current profile that can be characterized and used to estimate the moment when an operation completes.

Simple and differential power analysis (SPA and DPA) attacks are performed by monitoring device current consumption with very fine grained detail. These attacks seek to undermine encryption techniques by monitoring the current consumption of the processor and detecting the moment at which the processor executes a branch operation [10]. The attacks have been performed on an ARM Cortex MCU using AES and required an extensive measurement setup to accomplish [11]. PACER is inspired by this previous work using fine-grained in-circuit current measurement and fortunately benefits from much more lenient sampling requirements.

## III. METHODS

### A. Timing Heuristic PACER-T

Some peripheral operations exhibit highly deterministic timing qualities. Such a device is likely to be internally clocked and the operation is waiting for some number of clock cycles to expire before signaling that the operation completed. Such operations are typified Figure 3 in that neither the total energy consumed, nor the profile of that consumption are necessary to predict completion. Regardless of the power profile, the operation always completes within a narrow window of time. Erases and write operations to EEPROM and flash are typical examples of this behavior.

PACER-T uses the successive approximation algorithm shown in (4) to determine the optimal delay for an operation. The algorithm begins by executing an operation with the amount of delay specified in the device datasheet. After each iteration, if the operation completed earlier than predicted (Pass), then the amount of delay is halved. Otherwise, the operation was ongoing (Fail) and the next delay is increased by half the distance to the last previously successful operation.

$$Pass: \begin{cases} T_{upper} = T_{lower} \\ T_{lower} = T_{lower} - \frac{(T_{upper} - T_{lower})}{2} \end{cases}$$
$$Fail: \begin{cases} T_{lower} = T_{lower} + \frac{(T_{upper} - T_{lower})}{2} \end{cases} \quad (4)$$
$$Initial\ Conditions: T_{upper} = T_{worst-case}, T_{lower} = 0$$

The algorithm is executed online and provides the tightest possible timing. Upon expiration of the predicted wait period, if the device status register indicates that the operation is still occurring, then the algorithm has yielded an early prediction and it is appropriate to continue to wait. This would be considered the 'Fail' case of (4) and future estimates are increased. Otherwise, if the device status register indicates that the operation is complete, then the algorithm has yielded a late prediction and it is appropriate to reduce future estimates.

### B. Energy Heuristic PACER-E

Operations that consume a deterministic amount of energy are better characterized by PACER-E. For example, the operation might involve the charging of a storage element such as an inductor or capacitor. In any case, a certain amount of energy is required to complete the operation and once that energy requirement has been satisfied, the peripheral device considers the operation to be complete. Figure 4 is an example of an energy bound operation.

The energy based heuristic was performed similarly to PACER-T in that successive approximation is used. The system multiply-accumulates voltage and current samples fed to the peripheral device. When the digital integration has reached the test value, the operation is 'complete' and checked for correctness. The mechanics of (4) are applied to PACER-E, except that all T limits are replaced with E energy limits. PACER-E is slightly less precise than the timing based algorithm due to the time required to both sample and perform the digital integration necessary for threshold checking.

The energy consumed throughout a test is calculated using the fundamental relationship shown in (5). The results were calculated offline via (6) and (7), where S is the state of the device, and $T_s$ is the sampling period.

$$P = VI = \frac{E}{t} \quad (5)$$

$$E_s = \sum_{n=0}^{N-1} V_n I_n T_s \quad (6)$$

$$E_{total} = \sum_{S_0}^{S_{n-1}} E_s \quad (7)$$

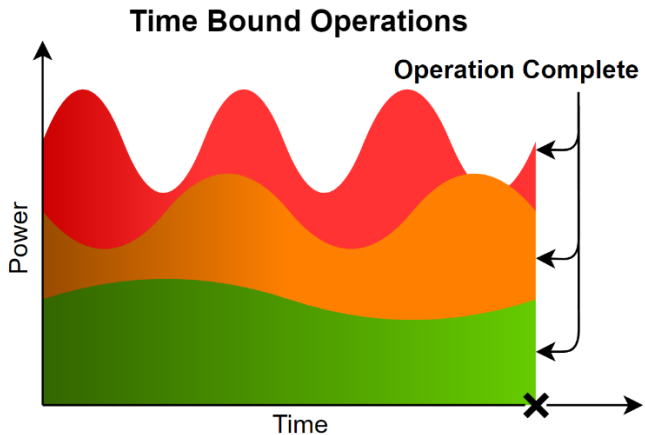

*Figure 3: Profile of Three Time-Deterministic Operations*

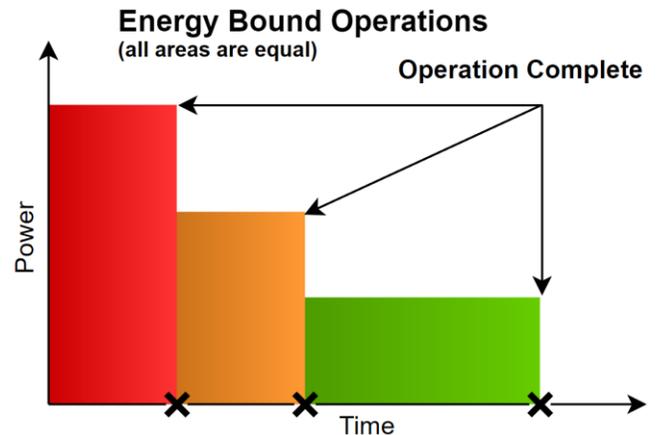

*Figure 4: Profile of Three Energy-Deterministic Operations*

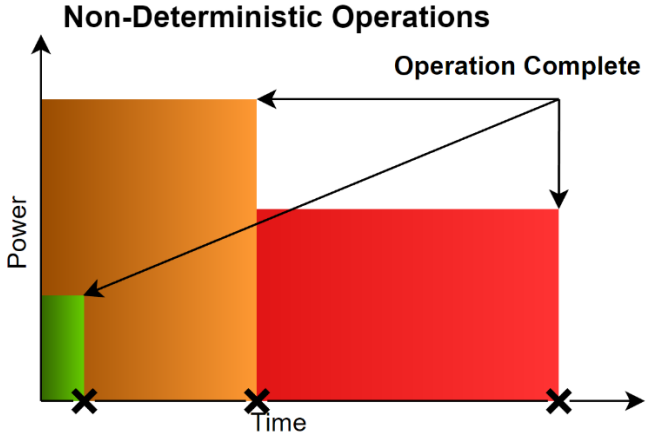

*Figure 5: Profile of Three Non-Deterministic Operations*

## C. Current Heuristic PACER-C

Some operations cannot reliably be defined in terms of time nor energy. One example of a non-deterministic operation would be communications tasks performed by Ethernet or wireless devices that have non-deterministic transmission latencies. Another example would be memory devices that incorporate an onboard memory hierarchy. In such devices, operations are affected by cache latencies.

PACER-C provides recognition that the operation is complete by measuring the idle current usage of the device before the operation begins and marking the operation as complete after the current returns to idle. In order to accommodate operations where the current returns to idle and yet the operation has not yet completed, the algorithm incorporates both a minimum latency and an idle current percent threshold to mark the operation as complete.

*Algorithm* 1: *PACER-C*

1:  ICT = (Idle Current Measurement) * threshold
2:  Execute Peripheral Device Operation
2:  **While (**t < Minimum Latency) and (I > ICT) **then**
3:     I = New Current Measurement
4:  **End While**

PACER-C is described by Algorithm 1 and begins by taking a sample of the device input current while idle. Next, the operation is executed and the algorithm waits for a minimum latency period to expire. The operation is considered complete after the output current returns to the threshold percentage of its previous state. The threshold for all following experiments was set empirically at 110%.

PACER-C is the most basic method to determine in real time if an operation has completed and may also be prone to false positives in some cases. There are many more advanced algorithms that can suit the purpose such as a multi-layer perceptron that is used in neural networks that could be used to identify features in real-time. It is notable however, that reducing the complexity of the detector is important so that the algorithm can ensure that it is keeping pace with incoming samples. Naturally, more complex algorithms could be accommodated by a more powerful host microcontroller.

## IV. MATERIALS

PACER and IODVS are implemented on an STM32F429 MCU supported by the STMicroelectronics DISCO board and hosted by the PRIME assembly. The board provides 64MB of SDRAM which allows for simultaneous sampling throughout the test suite at very high speed. All experiments were sampled at 1MSPS and the SDRAM allowed any individual experiment to last up to 1 full second. All of the analog conversions as well as the device state sampling were performed via DMA. Therefore, the test fixture is expected to have had no impact on the operation under test.

The PRIME (Precise Real-Time In-Circuit Micro-EMS) board, shown in Figure 6, hosts a variety of peripherals (labelled in red as DUT: Devices Under Test) that are commonly implemented in embedded designs. The board provides access to Bluetooth, Wi-Fi and a Si1143 proximity detector. PACER was evaluated on NAND and NOR FLASH memories, as well as a commercial EEPROM, temperature / humidity sensor and four independent Micro-SD cards.

At 1 MSPS and 4 channel measurements and 2 bytes per sample, each test can result in up to 8 megabytes of data. Because repeatability is so important, each test was run 50 times. Therefore, bandwidth became a limiting factor and a Hi-Speed (480Mbps) USB module was added to the board to allow for rapid development. Operating as a virtual communications port and using MCU parallel bus, actual bandwidth was realized at approximately 120Mbps.

Each of the peripheral devices under test has some method of determining if an operation completed successfully. For the memory devices, a simple read-back verification is sufficient to determine correctness and is a common practice among embedded designs. The temperature and humidity sensor provides a status bit indicating if an operation is in progress, thus indicating that a requested operation has not yet completed.

Power is provided and voltage is modulated to each individual device on the domain using independently configurable power supplies. The ASDM-300F module shown in orange on Figure 6 provides a high-efficiency buck power supply, followed by a linear regulator with a high ripple-rejection ratio. A high-precision and clean power supply is important because PACER uses the current profile to make real-time decisions. If the power supply outputs a significant amount of noise, then it becomes difficult to acquire signal and determine activity completion in real-time.

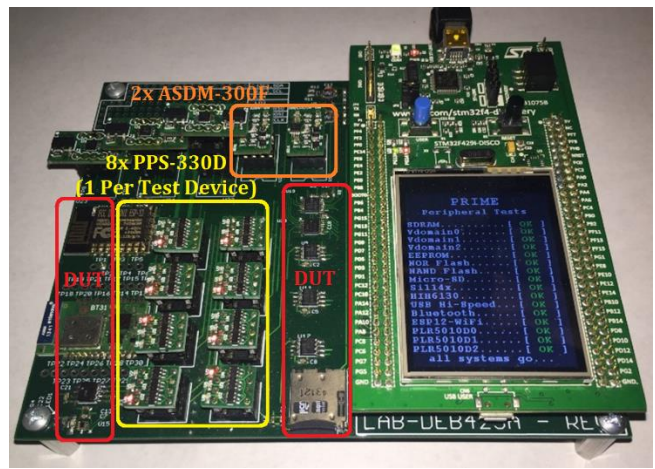

*Figure 6: PRIME (Precise Real-Time In-Circuit Micro-EMS)*

The ASDM-300F is also outfitted with a dual current measurement circuit using the Maxim MAX4377HAUA+. This circuit allows the host to measure both the input and output current of the power supply with high analog bandwidth. Ultimately, these outputs are used to determine activity completion with the PACER-E and PACER-C algorithms. It is important to note the gain-bandwidth product of the amplifier. High frequency content will be attenuated to some degree and the actionable data output would be of higher quality if a higher frequency device were available.

While measuring and classifying activity completion, it is important that each device be analyzed independently. The PPS-330D shown in yellow on Figure 6 allows the host to switch the voltage domain of an individual peripheral to any one of three domains, or disconnect the device entirely

PPS-330D devices are connected to each peripheral, and while a peripheral is under test, the remaining devices are switched to an alternate voltage domain. Thus, each device is independently classified in-system without physically removing other devices that may affect current measurements. Once the devices are characterized, then their individual contributions to the power supply current output can be deduced through superposition. The PPS-330D is convenient for initial profiling, but unnecessary for a streamlined implementation. The ASDM-300F is necessary for an IODVS implementation, but PACER-E and PACER-C only require the current measurement component.

## V. RESULTS

Initial IODVS results were repeated so as to establish a baseline with which to compare the results of PACER. Previous experiments required the results to be averaged many times over. The PRIME assembly provides high enough signal to noise ratio that averaging multiple test results is unnecessary and a simple 50-sample moving average provides enough filtering while maintaining a quick response time.

### A. MCP25AA512 EEPROM

The Microchip EEPROM is specified by the manufacturer for a 5ms mandatory wait period following the write command and data. This operation is highly deterministic with respect to time, energy and current profile. All PACER algorithms identified activity completion with high accuracy.

TABLE I.  MCP25AA512 EEPROM PACER RESULTS

| Stage | Latency Results (ms) | | | | |
|---|---|---|---|---|---|
| | *Control* | *PACER-T* | *Diff.* | *PACER+IODVS* | *Diff.* |
| Wait | 5.05 | 3.51 | 30.5% | 3.51 | 30.5% |
| All | 5.98 | 4.44 | 25.7% | 4.44 | 25.7% |
| | Energy Results (uJ) | | | | |
| Wait | 46.84 | 37.89 | 19.1% | 27.85 | 40.5% |
| All | 53.05 | 43.91 | 17.2% | 32.40 | 38.9% |

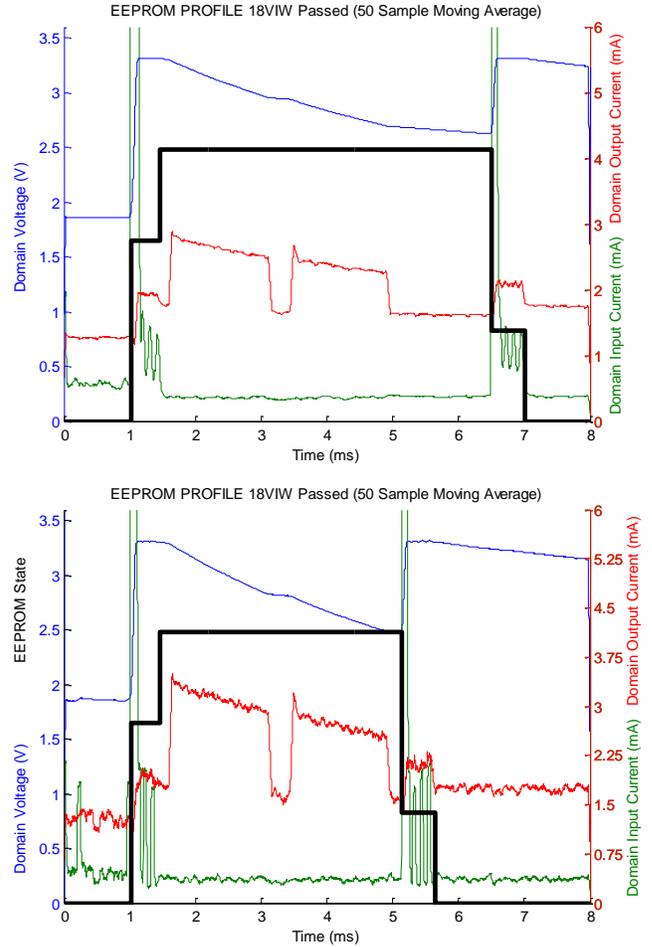

*Figure 7: EEPROM Write Cycle Using IODVS and PACER-T*

The current waveform of Figure 7 shows that the EEPROM write operation begins at t=1.5ms and indicates completion at approximately t=5ms instead of t=6.5ms as specified by the manufacturer. After applying the PACER-T algorithm, it is indeed true that the operation was complete at the 5ms mark, thus reducing the wait latency by 30.5%.

The PACER-E and PACER-C algorithms were also successful in identifying activity completion. The two algorithms do require additional computation to integrate or otherwise observe the current waveform. Given identical performance, PACER-T is the best choice in this application.

### B. Numonyx M25PX16 NOR Serial Flash

NOR flash modules sacrifice byte-wise modification for overall capacity. The M25PX16 presents 16MBits of capacity in a small package, but the host must erase sub-sectors of flash (4K) to write pages of flash (128B). To perform a read-modify-write operation, the host must read the contents of a sub-sector, modify the contents locally, erase the sub-sector in flash and finally write the modified contents back to the flash on a page-by-page basis.

Both the sub-sector erase and page write have a worst-case delay specified by the manufacturer. PACER algorithms were run against both operations to find the comprehensive result.

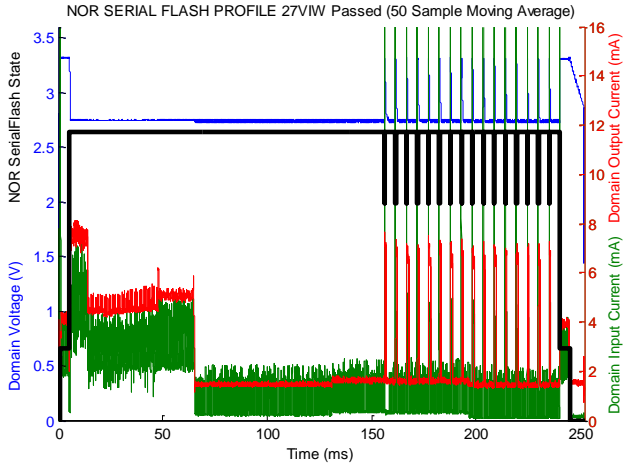
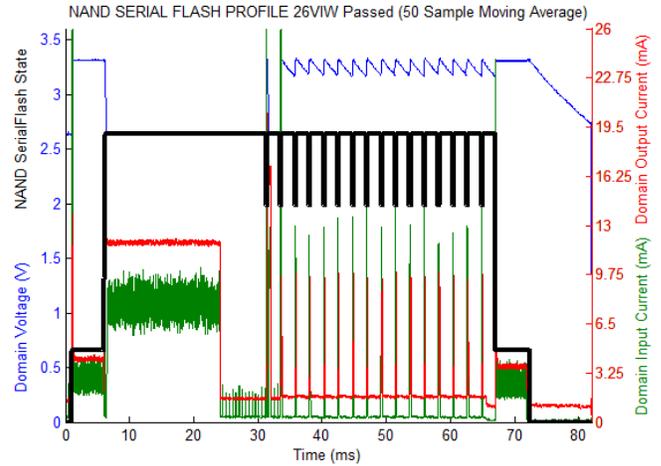
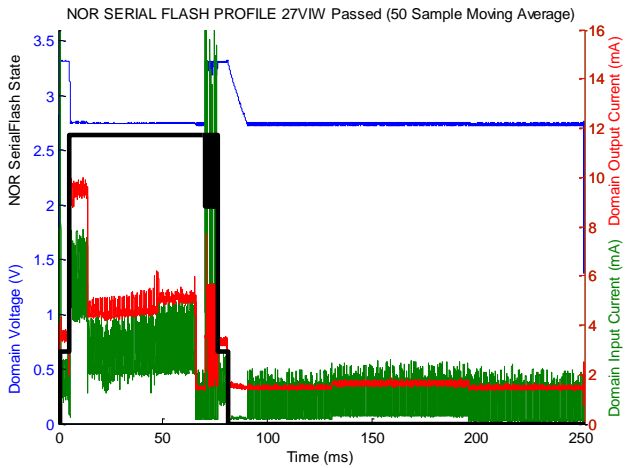
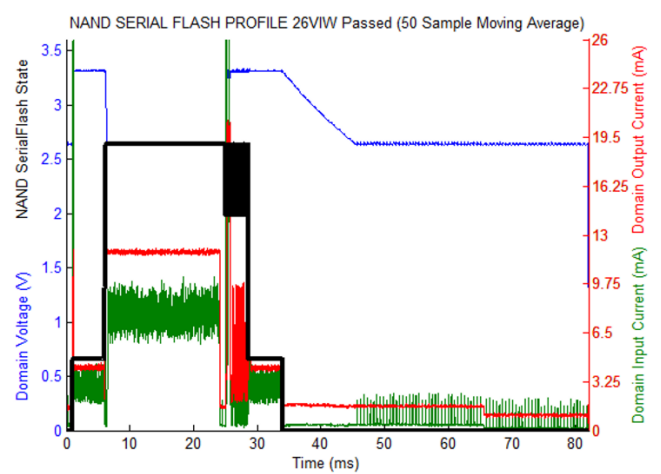

Figure 8: NOR Serial Flash Write-Cycle using IODVS and PACER-T

Figure 9: NAND Serial Flash Write Cycle Using IODVS and PACER-T

Although specified for 150ms, the current waveform indicates that the sub-sector erase completed approximately 65ms after it begins. Page writes are specified for a worst-case completion time of 10ms but through the application of PACER-T, they complete much faster as shown in TABLE II. The wait figure is the total amount of time spent waiting for the erase and the aggregate amount of time for each page write. The PACER-T algorithm delivered a 70% decrease in wait latency which yielded a 38.9% decrease in overall energy consumption. The worst-case manufacturer specification appears to be very pessimistic, although may be appropriate across both process and temperature variables.

### C. Microchip SST26VF016B Serial NAND Flash

The SST26 serial flash module uses NAND-like control logic to provide higher capacity and lower latency than the NOR serial flash. However, the device sacrifices the random-access timing benefit of NOR flash. The serial flash module must therefore read an entire page of flash into a local buffer before providing read data to the host. This can result in non-deterministic read and write access times.

While the core logic differs from the M25PX16, the PACER-T algorithm still performed the best. Application yielded a 66.6% decrease in aggregate wait times and a 17.8% decrease in overall energy consumption.

TABLE II.  M25PX16 NOR SERIAL FLASH PACER RESULTS

| Stage | Latency Results (ms) | | | | |
|---|---|---|---|---|---|
| | *Control* | *PACER-T* | *Diff.* | *PACER+IODVS* | *Diff.* |
| Wait | 231.57 | 69.47 | 70.0% | 66.92 | 71.1% |
| All | 243.87 | 82.45 | 66.2% | 80.26 | 67.1% |
| | **Energy Results (uJ)** | | | | |
| Wait | 2138.3 | 1212.0 | 43.3% | 1029.52 | 51.9% |
| All | 2277.0 | 1392.0 | 38.9% | 1158.26 | 49.1% |

TABLE III.  SST26VF016B NAND SERIAL FLASH PACER RESULTS

| Stage | Latency Results (ms) | | | | |
|---|---|---|---|---|---|
| | *Control* | *PACER-T* | *Diff.* | *PACER+IODVS* | *Diff.* |
| Wait | 57.61 | 19.26 | 66.6% | 19.27 | 66.6% |
| All | 71.28 | 32.94 | 53.8% | 32.95 | 53.8% |
| | **Energy Results (uJ)** | | | | |
| Wait | 1053.0 | 806.2 | 23.8% | 584.87 | 44.5% |
| All | 1247.9 | 997.26 | 17.8% | 801.95 | 35.7% |

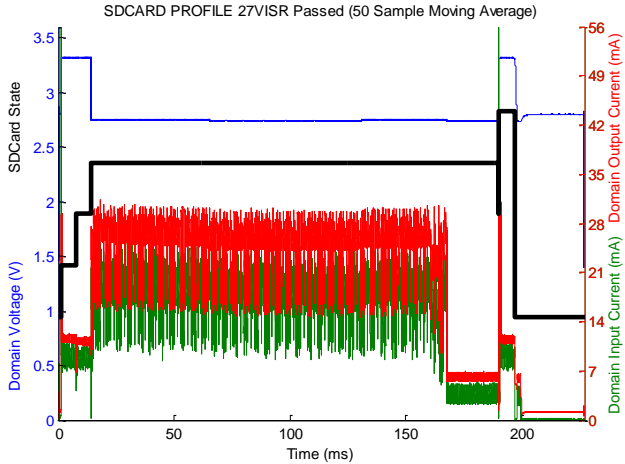

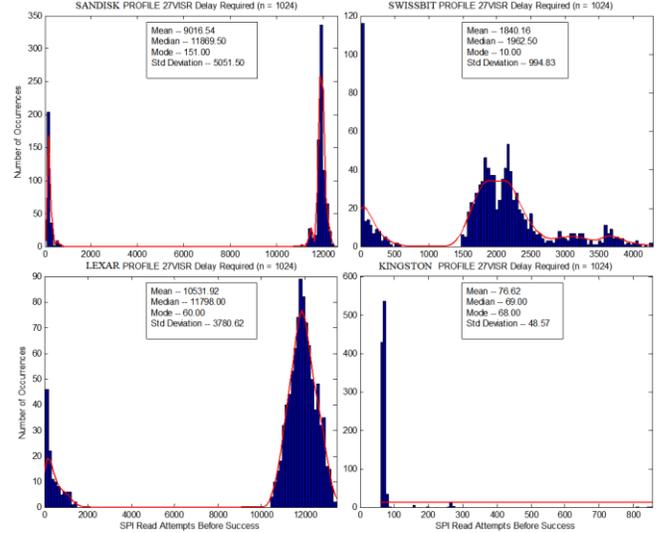

Figure 11: Timing Performance among Tested SD-Cards

TABLE IV. MICRO-SD CARD PACER RESULTS

| Stage | Energy Results (uJ) | | | | |
|---|---|---|---|---|---|
| | Control | PACER-C | Diff. | PACER+IODVS | Diff. |
| Sandisk | 17066 | 15198 | 10.9% | 11848 | 30.6% |
| Lexar | 22707 | 21428 | 5.6% | 16977 | 25.24 |
| Swissbit | 2763 | 914 | 66.9% | 554 | 80.0% |
| Kingston | 942 | 933 | 0.9% | 897 | 4.8% |

Figure 10: A Micro-SD Card Cache Miss and a Cache Hit

*D. An assortment of Micro-SD Memory Cards*

Onboard caches and memory management units cause the write operation of Micro-SD cards to have non-deterministic timing. In this case, PACER-C is the only algorithm that can reliably detect when the operation is finished. As with all memory tests, writes were performed with random data to random addresses throughout the memory space and so the cache performance is thoroughly exercised. Figure 10 shows the massive power and latency difference between a cache miss and a cache hit.

Figure 11 helps to describe the performance differences shown in TABLE IV. The control delay is set to the median delay for each characterization, PACER-C allows the host to react to those operations deviating considerably from the median. Therefore, the Sandisk and Lexar cards benefitted considerably because they exhibit a bimodal timing distribution. The Swissbit card benefits decisively because of the mostly normal timing distribution. The Kingston card does not benefit as much because write timing exhibits a very low standard deviation. To present complete timing effects, a thorough latency analysis would need to be done on each device. Only energy results are presented here, but they are correlated with overall latency decreases.

*E. Honeywell HIH-6130 Temperature / Humidity Sensor*

The Honeywell HIH-6130 communicates via the I$^2$C bus. The host requests the sensor to take a measurement and then waits the manufacturer-specified 45ms for the measurement to complete. Finally, the host retrieves the completed measurement. PACER-E demonstrated the best performance among the algorithms, perhaps because of the capacitive nature of the peripheral ADC. The effects are shown in Figure 12 and the numeric results are presented in TABLE V.

The PACER-T algorithm also produced impressive results with a wait latency of 31.66ms and wait energy of 254.14uJ. Compared with PACER-E, the result corresponds with a *slightly* increased latency and energy consumption of 0.5% and 4.3% respectively. For some applications, the simplicity of the PACER-T implementation may be preferable when compared to the best performing PACER-E algorithm.

TABLE V. HONEYWELL HIH-6130 PACER RESULTS

| Stage | Latency Results (ms) | | | | |
|---|---|---|---|---|---|
| | Control | PACER-E | Diff. | PACER+IODVS | Diff. |
| Wait | 45.27 | 31.45 | 66.6% | 19.27 | 66.6% |
| All | 45.99 | 32.17 | 53.8% | 32.95 | 53.8% |
| | Energy Results (uJ) | | | | |
| Wait | 325.95 | 240.29 | 26.3% | 169.62 | 48.0% |
| All | 330.50 | 245.39 | 25.8% | 173.89 | 47.4% |

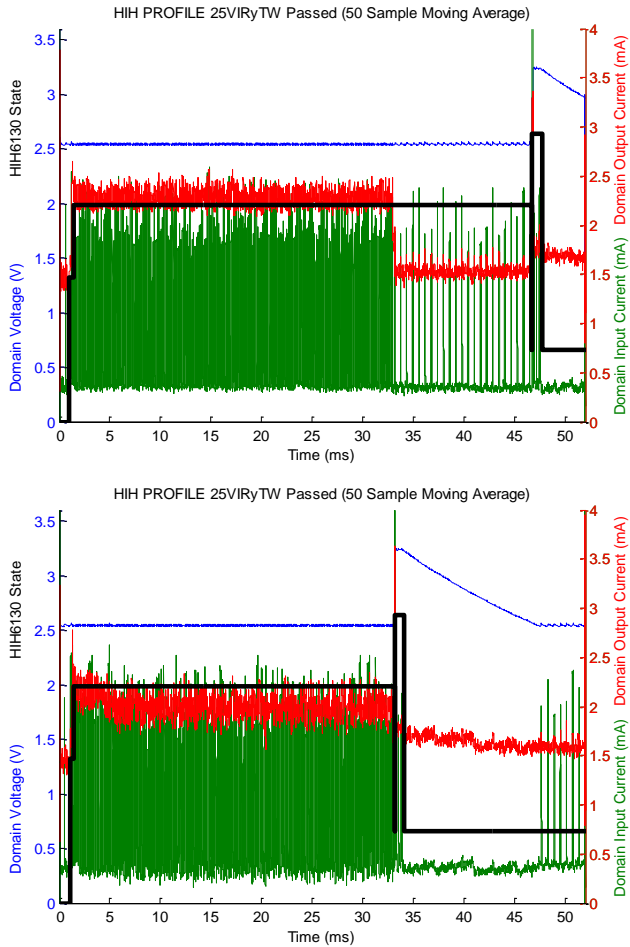

*Figure 12: HIH-6130 Measurement Cycle Using IODVS and PACER-E*

## VI. CONCLUSIONS

Applying the PACER suite of algorithms to a variety of common embedded peripherals resulted in significant reductions to both latency and energy consumption. The PACER-T algorithm performed best against time-bound operations and was very competitive in energy-bound operations. For non-deterministic operations, the PACER-C algorithm performed well. When measured against a median baseline, the algorithm performed even better as the operational latency increased in randomness.

The PACER-T and PACER-E algorithms use successive approximation and the PACER-C algorithm uses a return-to-idle measurement to determine activity completion. It is likely that the performance of both methods could be enhanced further through the application of more complex algorithms. PACER-T could be applied to memory operations with a bimodal delay distribution by first testing for a cache hit and then delaying for a determined cache-miss time. Likewise, the PACER-C algorithm could be modified online so as to identify the current waveform features corresponding varying latencies, thus allowing the MCU to sleep longer.

As the cost of computation in embedded systems continues to decrease, it is natural to devote more computational resources to minimizing system-wide energy consumption and latency. The PACER suite of algorithms use minimal computational resources and are shown to decrease latency by up to 67% and device energy consumption by up to 80% when compared to the naïve worst-case estimate.

## VII. REFERENCES

[1]   B. Brock and K. Rajamani, "Dynamic power management for embedded systems [SOC design]," in *SOC Conference, 2003. Proceedings. IEEE International [Systems-on-Chip]*, 2003.

[2]   C. Kumar, M. Sindhwani and T. Srikanthan, "Profile-based technique for Dynamic Power Management in embedded systems," in *Electronic Design, 2008. ICED 2008. International Conference on*, 2008.

[3]   W. Dargie, "Dynamic Power Management in Wireless Sensor Networks: State-of-the-Art," *IEEE Sensors Journal,* vol. 12, no. 5, pp. 1518 - 1528, 2012.

[4]   D. Moore and A. Dean, "Intra-Operation Dynamic Voltage Scaling," in *2015 IEEE 3rd International Conference on Cyber-Physical Systems, Networks, and Applications*, Hong Kong, 2015.

[5]   D. Lee, Y. Kim, G. Pekhimenko, S. Khan, V. Seshadri, K. Chang and O. Mutlu, "Adaptive-latency DRAM: Optimizing DRAM timing for the common-case," in *IEEE 21st International Symposium on High Performance Computer Architecture (HPCA)*, 2015.

[6]   S. Sadeghi-Kohan, M. Kamal, J. McNeil, P. Prinetto and Z. Navabi, "Online self adjusting progressive age monitoring of timing variations," in *10th International Conference on Design & Technology of Integrated Systems in Nanoscale Era (DTIS)*, 2015.

[7]   D. S. S. Etter, "Adaptive Estimatation of Time Delays in Sampled Data Systems," in *IEEE Transactions on Acoustics Speech and Signal Processing*, 1981.

[8]   S. G. P. A. Z. E. Tarasov V, "Efficient I/O Scheduling with Accurately Estimated Disk Drive Latencies," in *The Proceedings of OSPERT 2012*, 2012.

[9]   H. Macicior, M. Oyarbide, O. Miguel, I. Cantero, J. Canales and A. Etxeberria, "Iterative capacity estimation of LiFePO4 cell over the lifecycle based on SoC estimation correction," in *Electric Vehicle Symposium and Exhibition (EVS27)*, 2013.

[10]   H. Mahanta, A. Azad and A. Khan, "Power analysis attack: A vulnerability to smart card security," in *International Conference on Signal Processing And Communication Engineering Systems (SPACES)*, 2015.

[11]   M. Petrvalsky, M. Drutarovsky and M. Varchola, "Differential power analysis attack on ARM based AES implementation without explicit synchronization," in *Radioelektronika 2014 24th International Conference*, 2014.